\documentclass[floatfix,preprint,showpacs,preprintnumbers,amsmath,amssymb,superscriptaddress]{revtex4-1}
\usepackage{graphicx,pdfpages,color,soul}

\begin{document}



# Evidence for hydrodynamic electron flow in PdCoO$_2$

Philip J. W. Moll, [1,2,3] Pallavi Kushwaha, [3] Nabhanila Nandi, [3]
Burkhard Schmidt, [3] Andrew P. Mackenzie, [3,4*]

[1]Laboratory for Solid State Physics, ETH Zurich,
CH-8093 Zurich, Switzerland,

[2]Department of Physics, University of California,
Berkeley, California 94720, USA,

[3]Max Planck Institute for Chemical Physics of Solids,
Nöthnitzer Straße 40, 01187 Dresden, Germany

[4]Scottish Universities Physics Alliance, School of Physics and Astronomy,
University of St. Andrews, St. Andrews KY16 9SS, United Kingdom

*To whom correspondence should be addressed; E-mail: Andy.Mackenzie@cpfs.mpg.de

**Electron transport is conventionally determined by the momentum-relaxing scattering of electrons by the host solid and its excitations. Hydrodynamic fluid flow through channels, in contrast, is determined partly by the viscosity of the fluid, which is governed by momentum-conserving internal collisions. A long-standing question in the physics of solids has been whether the viscosity of the electron fluid plays an observable role in determining the resistance. Here we report experimental evidence that the resistance of restricted channels of the ultra-pure two-dimensional metal PdCoO$_2$ has a large viscous contribution. Comparison with theory allows an estimate of the electronic viscosity in the range between $6\times10^{-3}$ kg(ms)$^{-1}$ and $3\times10^{-4}$ kg(ms)$^{-1}$, to be compared with $1\times10^{-3}$ kg(ms)$^{-1}$ for water at room temperature.**

In a quantum fluid without an associated lattice, such as $^3$He, the momentum of the fluid is conserved except where it interacts with the walls of a channel through which it is flowing. As the temperature decreases and the quasiparticle-quasiparticle mean free path $\ell$ within the fluid increases because of the decrease of its quasiparticle scattering rate, interactions with the walls become more probable, and the viscosity and flow resistance increase. This is intuitively at odds with the behavior seen for electrons moving in a crystalline lattice, whose flow resistance decreases as $\ell$ increases. The resolution of this apparent paradox is that coupling to the lattice and its excitations means that the large majority of collisions in the electron fluid (electron-impurity, normal electron-phonon, Umklapp electron-electron and Umklapp electron-phonon) relax momentum, taking the fluid far from the hydrodynamic limit. At least some of these momentum-relaxing collisions are inevitable in any real material. Strictly speaking, momentum of the electron fluid can never be conserved, even in a bulk sample for which boundary scattering is insignificant. This does not, however, mean that the electronic viscosity needs to play no role in determining electrical resistance. A pragmatic benchmark is whether momentum-conserving processes are faster or slower than momentum-relaxing ones. If the electron fluid's momentum is relaxed slowly, it can be thought of as being quasi-conserved, and hydrodynamic signatures might be observable *(1-9)*.

The search for hydrodynamic effects in electrons in solids has been given extra impetus by the introduction of the "holographic correspondence" to condensed matter physics *(10)*. This technique introduced the concept of a minimum viscosity, argued to be applicable to strongly interacting fluids as diverse as the quark-gluon plasma and cold atomic gases *(11)*.

2

Hydrodynamic effects have also been postulated to be at the root of the *T*-linear resistivity of the high temperature superconductors *(6, 7)*, but because momentum-relaxing scattering is strong in those materials, it is difficult to perform an analysis of the experimental data that unambiguously separates the two effects. In a pioneering experiment, unusual current-voltage relationships in a semiconductor wire were convincingly ascribed to hydrodynamic effects *(3)*, but that avenue of research has not been widely pursued, even though the large difference between transport and electron-electron scattering rates in semiconductors was subsequently demonstrated by direct non-equilibrium measurements *(12)*.

Here we sought to identify a material in which momentum-relaxing scattering is anomalously suppressed in order to investigate whether a hydrodynamic contribution to electrical transport could be clearly separated from the more standard contributions from momentum-relaxing processes. The material that we chose was $PdCoO_2$, a layered metal with a series of unusual properties *(13- 21)*. Its electronic structure is deceptively simple, with one highly dispersive band, dominantly of Pd 4*d*/5*s* character, crossing the Fermi level *(22-26)*. Its Fermi volume corresponds to one electron per formula unit to high accuracy *(18)* and the ratio of in-plane to out-of-plane resistivity is approximately $10^3$, justifying the use of a two-dimensional approximation in treating the in-plane properties.

The electrical conductivity of $PdCoO_2$ is remarkable. At room temperature, its resistivity is just 2.6 $\mu\Omega$ cm, 30% lower per carrier than that of elemental copper. Below 15 K, the resistivity is essentially independent of temperature, and is below 10 n$\Omega$ cm in the best single crystals *(18)*. This striking behavior might be attributable to phonon drag, in which the phonons follow the electrons into an out-of-equilibrium distribution when an electric field is

3

applied. In PdCoO$_2$, the activation temperature for Umklapp electron-phonon processes is at least 160 K, unusually high for a metal *(18)*.

The above properties make PdCoO$_2$ a good candidate for a search for hydrodynamic effects. Below 15 K, momentum-relaxing processes are far slower than those observed in most metals. Better still, if phonon drag is indeed taking place, the normal electron-phonon processes that usually help prevent electrons in solids from approaching the hydrodynamic limit are now helping that process, because they contribute a source of momentum-conserving scattering. An important independent hint that the balance between momentum-conserving and momentum-relaxing scattering is both unusual and favourable for the observation of hydrodynamic effects comes from the ratio of the mean free paths deduced, respectively, from the resistivity, which is sensitive to processes that efficiently relax momentum, and from analysis of the de Haas-van Alphen effect, whose amplitude is sensitive to a wider range of scattering processes. The inverse of that ratio is 5-10 %, an unusually small value approximately an order of magnitude lower than that seen in ordinary metals *(18)*.

In a purely hydrodynamic fluid, flow resistance in channels is determined entirely by momentum-relaxing boundary scattering, the efficiency of which is determined by the fluid's viscosity. Boundary scattering also contributes to the resistance in thin wires or channels of metals in which hydrodynamic effects play no observable role, but the standard theory of this phenomenon is well established *(27)*. We therefore set out to construct a series of PdCoO$_2$ "wires", and study how their resistance varied with the channel width and investigate if there were observable deviations from the predictions of standard theory. The wires were produced from flux-grown single crystals using Focused Ion Beam (FIB) etching (Fig. 1A). Six such devices

were made and shown to have consistent properties *(28)*. We discuss the nature of the boundaries produced by FIB processing in *(28)* and show that damage is restricted to approximately 20 nm from the edge of the channel. Taking all uncertainties into account, the undamaged channel width can be determined to an accuracy of ± 80 nm. For the main experiment, measurements were done on a single successively etched wire from the same crystal, to remove as many experimental uncertainties as possible. A second crystal (Fig. 1B) was etched into a meander channel ideally suited for a measurement of the Shubnikhov-de Haas (SdH) effect. For each experiment, we studied the magnetoresistance of the wire in magnetic fields $B$ of -14 T $\leq B \leq$ 14 T. We also fabricated a multi-contact device used to verify that our data are length-independent at constant width *(28)*.

Data obtained in the channel narrowing experiment are shown in Fig. 2A, for widths ranging from 60 $\mu$m to 0.7 $\mu$m, at a measurement temperature of 2 K. Data from the meander channel are shown in Fig. 2B. Consistent with previous measurements on other single crystals *(18)*, the resistivity in zero field for the 60 $\mu$m wide sample is 0.009 $\mu\Omega$ cm. The value of the momentum-relaxing mean free path $\ell_{MR}$ is a crucial parameter in the analysis of the resistance of restricted channels; the single-band electronic structure and well-known Fermi surface shape and volume *(16, 18)* allow an accurate calculation of $\ell_{MR}$ =18.5+1.5 $\mu$m *(28)*. The wire widths $W$ used for the experiment therefore cover the range 0.3 $\leq \ell_{MR}/W \leq$ 26, enabling study of the crossover between a nearly bulk regime and one in which a sample dimension falls far below the bulk mean free path.

We adopt the usual solid state physics convention of describing the transport properties of our channels in terms of the resistivity $\rho$. For a channel of width $W$, length $L$ and thickness

$T$, $\rho \equiv RTW/L$. Conceptually, $\rho$ is a bulk property of the material, so in the absence of boundary effects it should be independent of $W$. In contrast, at low fields, the overall channel resistivity $\rho$ increases by over an order of magnitude as the wire width is decreased (Fig. 2). Because this involves repeated exposure to ion beam etching, it is natural to wonder whether this trend is caused by beam damage increasing the scattering in the bulk of the wire. However, extending the data to higher fields proves that this is not the case. Firstly, we note that, at high fields, the resistivity is similar at all widths above 0.7 $\mu$m. Secondly, the pronounced maximum seen at fields $B_{max}$ in the MR for $W \leq$ 30 µm is a well-known phenomenon from the study of narrow conducting channels for which the bulk mean free path is of the order of the channel width or larger *(29-31)*. For each channel width, the rise in the magneto-resistivity at low fields is stopped when the cyclotron orbit radius falls to less than the channel width, because the helical pitch of the motion of the drifting electrons becomes so tight that boundary scattering is suppressed, and eventually the bulk, width-independent resistivity is recovered at high fields. Specifically, $B_{max} = \alpha \frac{\hbar k_F}{We}$ (where $\hbar$ is Planck's constant divided by 2π, $k_F$ is the Fermi wave vector and *e* the electronic charge) with constant $\alpha$ = 0.55 has been reported for restricted channels of two-dimensional electron gas *(30)*, and a similar functional form with $\alpha$ = 0.9 observed in recent work on graphene *(31)*. The data shown in Fig. 2C therefore provide further evidence that the overall scale of the resistivity is increasing because of additional boundary rather than additional bulk scattering. Very low field peaks in the magnetoresistance can still be seen in the 30 µm wide channel, providing further direct geometrical evidence that $\ell_{MR}$ is long. Final evidence that extremely high bulk purity is retained after ion beam etching comes from the data from the meander sample. Its $B_{max}$ is

consistent with the measured width of 6 $\mu$m (Figs. 2, B and C) and at high fields, clear SdH oscillations are seen, with frequencies in agreement with those seen in dHvA from bulk crystals *(18)*. High frequencies are particularly hard to see by the SdH effect; the 30 kT frequencies shown in Fig. 3B are among the highest ever reported in Shubnikov-de Haas measurements.

The data presented in Fig. 2 and Fig. 3 give strong evidence that boundary rather than bulk scattering dominates the rise in resistivity seen as we reduce the channel width. As discussed above, boundary scattering is expected even in the absence of hydrodynamic effects as the channel width falls to less than the mean free path and the system enters the ballistic transport regime. The relevant theory *(4, 27)* can be expressed in a useful dimensionless form, shown in the blue line in Fig. 4. If $\rho$ is normalized to the bulk resistivity $\rho_o$ of an infinitely wide sample, and plotted against $\ell_{MR}/W$, the prediction has no free parameters. By $\ell_{MR}/W$ = 25, $\rho$ is calculated to be 10.3 $\rho_o$. Our measured value for $\rho/\rho_o$ is over 50% larger than this prediction, and our data also show a functional form that is at odds with the simple theory.

In order to examine whether or not the large deviations of the data from the predictions of standard transport theory are linked to electronic hydrodynamics, we have carefully studied the predictions of a more sophisticated theory that takes momentum-conserving scattering into account. Originally formulated to analyze current-dependent hydrodynamic signatures reported in Ref. *(3)*, the theory encodes momentum-relaxing scattering via the role of impurities and momentum-conserving scattering via normal electron-electron scattering processes. In PdCoO$_2$, the scale of momentum-conserving electron-electron scattering is uncertain due to Fermi surface faceting *(26)* and phonons dragged out of equilibrium are

likely also to be making a contribution to the momentum-conserving processes. In this sense, even this more sophisticated theory is over-simplified, and it would be dangerous to use it to predict temperature-dependent transport in PdCoO$_2$. However, its results at any fixed temperature depend only on the ratio of a momentum-conserving mean free path $\ell_{MC}$ to the momentum relaxing mean free path $\ell_{MR}$, and not, in detail, on the microscopic origin of the scattering that produced that ratio. It is therefore a useful guide to the consequences of including hydrodynamic effects in experiments performed at constant temperature such as the one summarized in Fig. 4, A and B. For any value of $\ell_{MC}/\ell_{MR}$ it predicts a unique functional form and overall magnitude for $\rho/\rho_o$ vs $\ell_{MR}/W$, with no free fitting parameters.

As can be seen from the red lines in Fig. 4, A and B, the hydrodynamic theory for $\ell_{MC}/\ell_{MR}$ = 0.1 produces an excellent match to our data. In Fig. 4C we show how the predictions of the theory change as a function of $\ell_{MC}/\ell_{MR}$, marking the blue and red lines of Figs. 4, A and B on the contour plot for reference. For large $\ell_{MC}/\ell_{MR}$ (i.e. weak momentum-conserving scattering) it limits rapidly to the prediction of the standard theory. At its other limit of strong momentum-conserving scattering (black line) the prediction for $\rho/\rho_o$ is approximately quadratic as a function of $\ell_{MR}/W$. In this regime, viscous effects dominate, and the $W^{-2}$ dependence of $\rho$ is equivalent to the $W^{-3}$ prediction for flow resistance that is obtained from the purely hydrodynamic Navier-Stokes equation *(28)*.

Our value for $\ell_{MC}/\ell_{MR}$ (red line) sits between these limiting cases. Since viscosity is inversely proportional to $\ell_{MC}$, the initial $\sim W^{-2}$ rise is steeper than for $\ell_{MC}/\ell_{MR}$ = 0.01. As $\ell_{MR}/W$ rises towards 10, $\ell_{MC}$ is no longer much less than $W$, and the Navier-Stokes prediction evolves smoothly to a solution in which viscous effects are important but the

channel constriction is such that we leave the purely hydrodynamic regime. Physically the system is in a hybrid situation in which traditional impurity and boundary scattering mix with viscosity-stimulated boundary scattering to produce the overall evolution of resistivity with channel width. Fig. 4 C also shows that the hydrodynamic prediction is insensitive to the precise choice of $\ell_{MC}/\ell_{MR}$. Our choice of 0.1 was not the result of fitting, but simply an estimate motivated by the ratio of scattering rates deduced from measurements of resistivity and dHvA. Choices a factor of two larger or smaller would give a similar level of agreement with the data.

The data and predictions shown in Fig. 4 and discussed further in *(28)* provide strong evidence that we have observed a substantial hydrodynamic contribution to electrical transport in a bulk material. Further analysis of the theory in its Navier-Stokes limit allows a quantitative estimate of the electronic viscosity itself as a function of the hydrodynamic contribution to $\rho/\rho_0$. As a result, we are able to estimate the dynamic viscosity $\eta$ of the electronic fluid in PdCoO$_2$ as lying in the range between $6\times10^{-3}$ kg(ms)$^{-1}$ and $3\times10^{-4}$ kg(ms)$^{-1}$. For comparison, those of two well-known fluids, water at room temperature and liquid nitrogen at 75 K, are $1\times10^{-3}$ kg(ms)$^{-1}$ and $1\times10^{-4}$ kg(ms)$^{-1}$ respectively.

Although the bounds we can place on the electronic viscosity of PdCoO$_2$ are not very precise, we stress that this is an issue of theory, not experiment. In principle, we believe that there is such a large hydrodynamic contribution to our data that they contain all the information required to estimate the viscosity precisely, and hope that our experiment motivates further work on this issue. As discussed throughout the paper, PdCoO$_2$ is a weak-scattering system, so it is not surprising that comparing our estimate of $\eta$ with

measurements of the entropy density $s$ (18, 15) yields $\eta/s \sim 10^6\ \hbar/k_B$, far from a proposed minimum viscocity limit (11). It will also be interesting to re-examine a possible role of hydrodynamic effects in explaining the resistivity in systems in which the momentum-conserving scattering is extremely strong (6,7,32). In principle, a range of viscosities is to be expected in different electronic fluids; turbulent electronic flow might even be attainable in future.

Supplementary Materials

Materials and Methods
Figs. S1 to S6
References 33-39

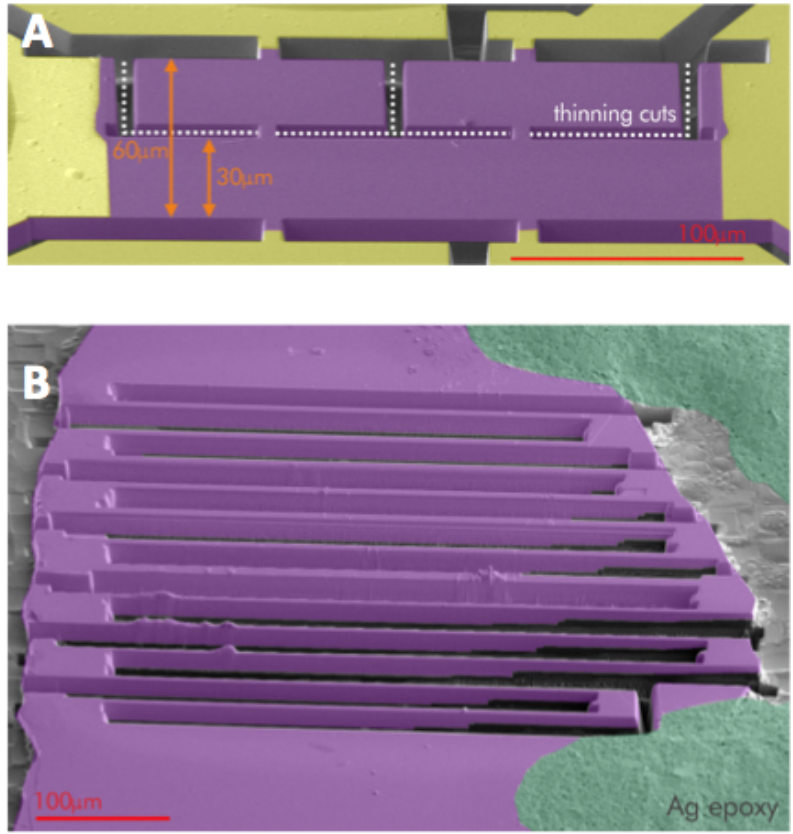

**Fig. 1. FIB prepared devices of PdCoO$_2$ crystals.** (A) The crystal used for our channel thinning experiment, after the first two processing steps. An initial channel 120 $\mu$m wide has been re-processed to produce a narrower conducting channel of width 60 $\mu$m. The same channel was subsequently reprocessed seven further times, narrowing it in approximately factor of two steps until it was 0.7 $\mu$m wide. (B) A meander channel processed in a second crystal for use in a search for the Shubnikhov-de Haas effect. Its approximate width is 6 $\mu$m in the narrowed parts and thickness is 17 $\mu$m. Its resistance in a magnetic field of 14 T at a temperature of 2 K is 5.7 $m\Omega$.

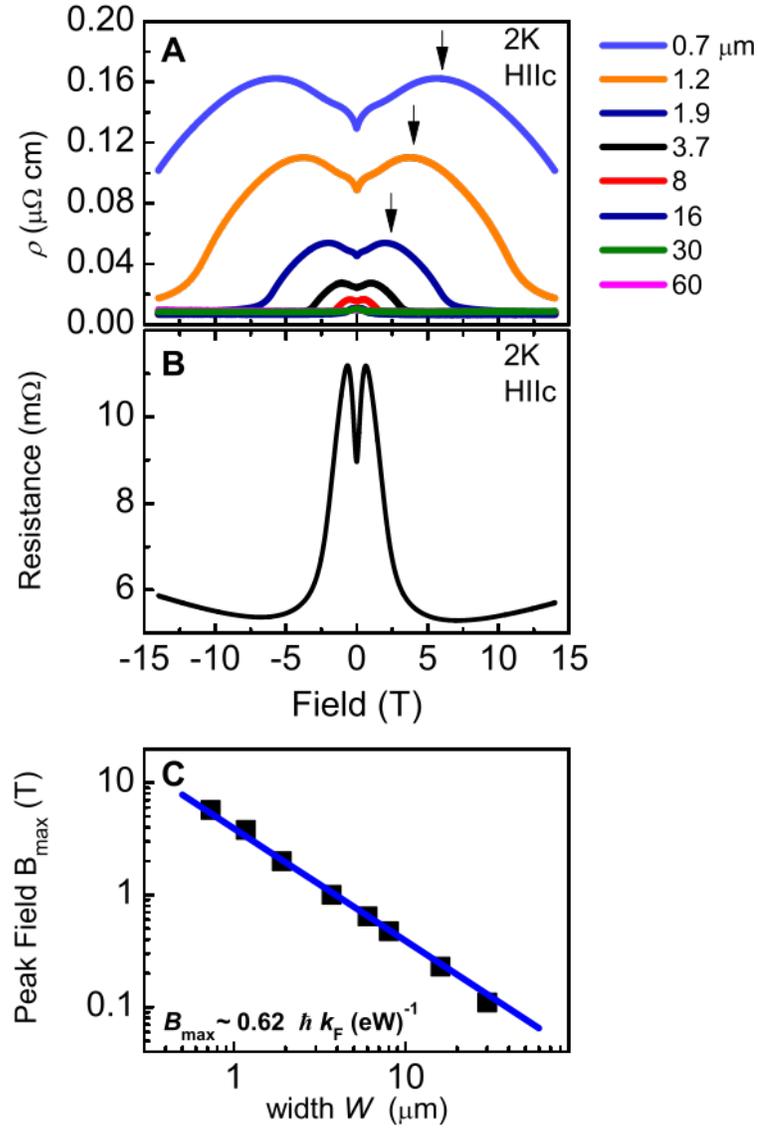

**Fig. 2. Effect of channel width (*W*) on magnetotransport.** (A) Magnetoresistance data from the samples shown in Fig. 1A taken at 2 K after successive channel thinning steps. (B) Magnetoresistance data from the sample shown in Fig. 1B taken at 2 K. (C) $B_{max}$ (indicated by arrows in (A)). varies as $\alpha \frac{\hbar k_F}{We}$ as the device shown in Fig. 1A is successively thinned. Both the functional form and prefactor are in good agreement with theory and with previous measurements on semiconductor wires.

15

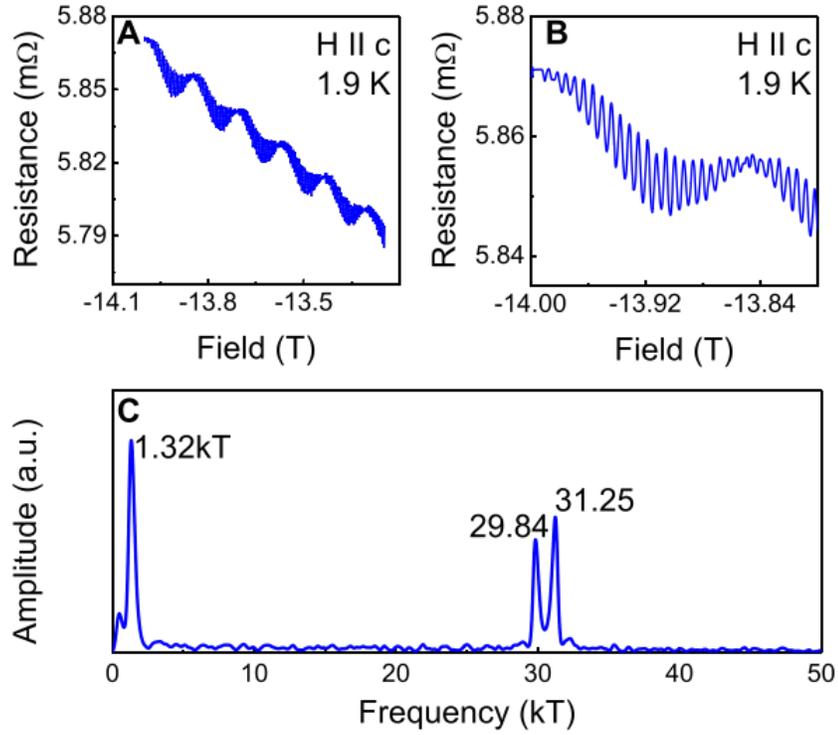

**Fig. 3. Shubnikhov-de Haas oscillations.** (A, B) Shubnikhov-de Haas oscillations from the patterned meander track shown in Fig. 1B . (C) The frequencies of the SdH oscillations extracted by Fourier analysis of the data in (A).

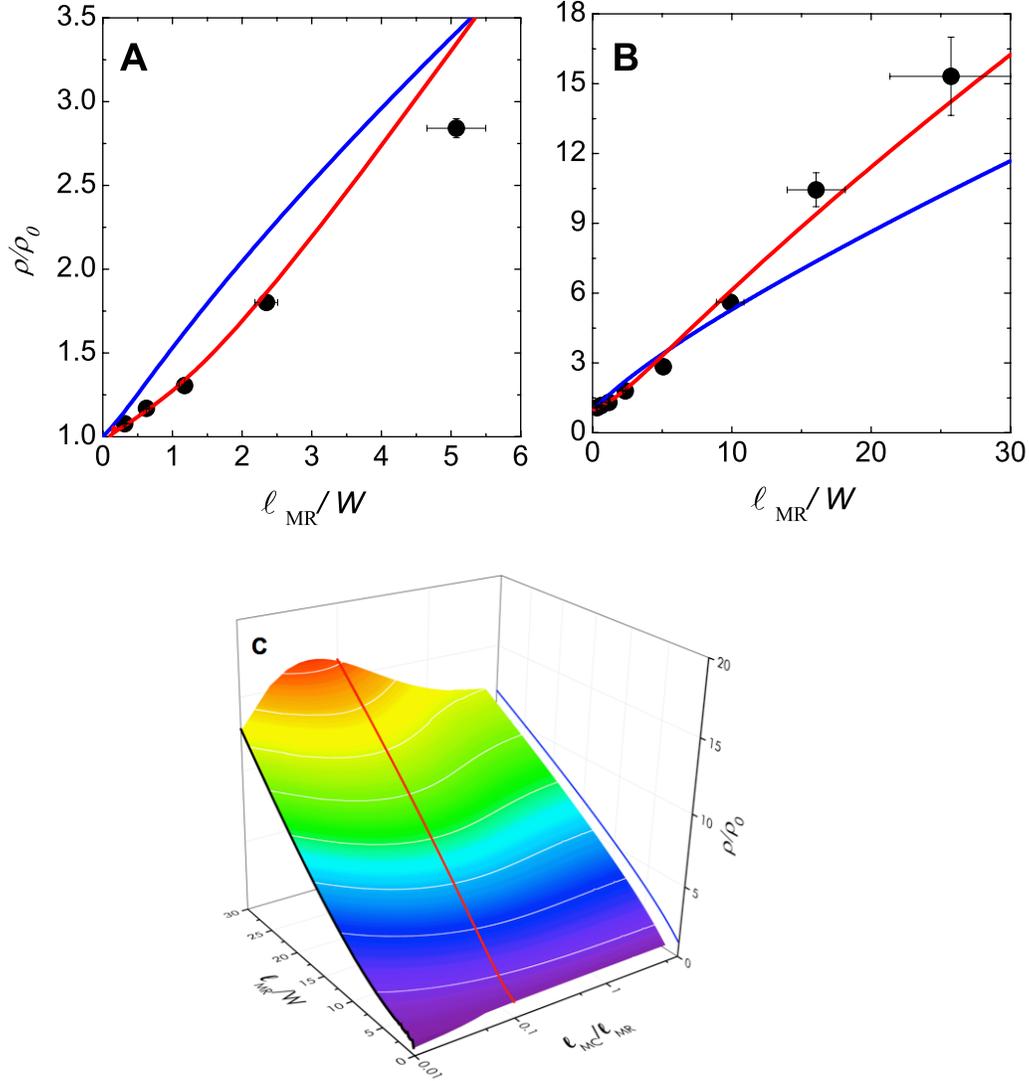

**Fig. 4. Hydrodynamic effect on transport.** (A, B) The measured resistivity of PdCoO$_2$ channels normalised to that of the widest channel ($\rho_0$), plotted against the inverse channel width $1/W$ multiplied by the bulk momentum- relaxing mean free path $\ell_{MR}$ (closed black circles). Blue solid line: prediction of a standard Boltzmann theory including boundary scattering but neglecting momentum-conserving collisions (Red line: prediction of a model that includes the effects of momentum-conserving scattering (see text). In (C) we show the predictions of the hydrodynamic theory over a wide range of parameter space.

17



\setcounter{equation}{0}
\setcounter{figure}{0}
\setcounter{table}{0}
\setcounter{page}{1}
\makeatletter
\renewcommand{\theequation}{S\arabic{equation}}
\renewcommand{\thefigure}{S\arabic{figure}}
\renewcommand{\bibnumfmt}[1]{[S#1]}
\renewcommand{\citenumfont}[1]{S#1}
\makeatother

\textbf{MATERIALS AND METHODS}

\section{CRYSTAL GROWTH AND EXPERIMENTAL METHODS}

\subsection{Crystal growth and characterisation.}

Single crystals were grown in sealed quartz tubes, using the reaction PdCl$_{2}$ + 2CoO $\rightarrow$ PdCoO$_{2}$ + CoCl$_{2}$, and extracted in hot ethanol.  We broadly followed procedures described in Ref.~\textit{(17)} , but experimented with modified temperature profiles to optimize the growth.  Standard x-ray diffraction, chemical analyses and transport measurements were used for initial characterization of the phase purity and high conductivity of the crystals.

\subsection{Calculating resistivity and the momentum-relaxing mean free path. }

In a rectangular parallelepiped with homogeneous current flow, the resistivity ρ is deduced from measured resistance using the formula $\rho$ = $RLW/T$, where $R$ is the measured resistance, $L$ the spacing between the voltage contacts, $W$ the width and $T$ the physical thickness of the sample.  In a highly two-dimensional material such as PdCoO$_{2}$, whose in-plane conductivity is a thousand times larger than that between the planes, the effective electrical thickness $T_e$ can differ from $T$: if the voltage contacts are not far enough from the current injection point, the current has not spread through the entire thickness of the crystal before passing between the voltage contacts.  In bulk measurements this issue is typically avoided by attaching current leads over the end faces of the sample in an attempt to inject the current evenly across the whole thickness. In this work, however, the sample and contact geometries involved current injection through the top surface.  Calculation for a parallelepiped modeled on the device shown in Fig.~1A  showed that a modest difference between $T$ and $T_e$ was likely.  The main resistivity results are normalized in such a way that any correction factor between $T$ and $T_e$ would drop out from the $\rho/\rho_{0}$ axes of Fig.~4, but the factor is important to accurate determination of the momentum-relaxing mean free path $\ell_{MR}$.  Rather than rely quantitatively on such a calculation, which would inevitably involve some assumptions, we fabricated the extra device shown in Fig.~S1, from a crystal of similar thickness to that shown in Fig.~1 A.  By studying the evolution of the measured $\rho$ along the length of this multi-contact device, we determined that, for a device with the geometry shown in Fig.~1 A, $T_e$ =$ \alpha T$, with $\alpha$ = 0.75 $\pm$ 0.05.  Analysis of our channel data using $\alpha$ = 0.75 and allowed determination of $\rho_0$ = 8.5 $\pm$ 0.6 n$\Omega$cm. This is in excellent agreement with other measurements on bulk crystals from the same batch.

The error associated with converting the measured resistivity $\rho_0$ to the bulk mean free path is lower than that in determining $\rho_0$.  In a two-dimensional metal, the resistivity is given by the line integral of the mean free path $\ell$ around the Fermi ``surface''.  At the low temperatures relevant to the data shown in Figs. 2~-~4, $\ell$ can be assumed to be independent of k in the isotropic-$\ell$ approximation~\textit{(33)} but in P\lowercase{d}C\lowercase{o}O$_{\text2}$ this is a very mild approximation since the Fermi velocity v$_F$ is almost k-independent in any case.  If the Fermi surface has circular cross-section we arrive at the famous two-dimensional expression $\ell$ = $\frac{\textit{hd}}{{\textit{e}^2\textit{k}_F\rho}}$, where $\textit{k}_F$ is the Fermi wave vector and $\textit{d}$ is the interlayer spacing. 

If the Fermi surface is not circular, a correction is required to this formula because, for a given area, the perimeter around which the line integral is performed is longer than that of a circle. The correction is usually small for the shape close to a circle. For the rounded hexagonal Fermi surface of PdCoO$_2$ ~\textit{(16, 25)}  it is straightforward to estimate it numerically, and it produces a 2$\%$ change in the calculated mean free path from that estimated by simply using Eq. S1 and k$_F$ defined as $(A/\pi)^{1/2}$ where the Fermi surface area $A$ is known to within 1$\%$ accuracy from the quantum oscillation frequencies. The perimeter change was taken into account in our calculation.  Taking into account the uncertainty in $A$, and combining with that in $\rho_0$ yields $\ell_{MR}$ = 18.5 $\pm$ 1.5$\mu m$.

After completing our channel narrowing experiments, we verified the validity of our two-dimensional approximation by etching the narrowest channel from above, to reduce its thickness.  Measurements before and after this step showed the expected change in resistance but no change in resistivity, as expected for a two-dimensional material. 

\subsection{Focused Ion Beam sample fabrication}

Using Focused Ion Beam (FIB) fabrication to study the width dependence of the resistance of micro-channels is a relatively new approach and thus special care must be taken to investigate potential issues arising from this technique. In particular, the nature of defect generation due to the 30kV ion irradiation in the crystal bulk as well as the sidewalls of the microwires need to be considered. In the following, we will discuss the fabrication details and the expected state of the sidewall surfaces. For a general introduction into the details of FIB micromachining, we refer to ~\textsl{(34)}  and~\textsl{(35)}.

{\bf FIB fabrication:} The PdCoO$_{2}$ crystals grow naturally as thin platelets, with a typical in-plane width in the few 100 $\mu$m range and a typical thickness around 10 $\mu$m. Crystals were first screened under an optical microscope for evident macroscopic defects such as strong terrace growth, cracks and intergrowths. Platelets with clear hexagonal morphology and immaculate surfaces were chosen for further fabrication. The crystals were glued with epoxy onto a silicon chip and sputter-coated with 150 nm of gold. This gold was structured by FIB milling into the desired contact configuration. A Helios Nano Lab 600i by FEI was used for the sample preparation.

In a following step, the crystals were coarsely structured into their final geometry. FIB milling is a fairly gentle way of structuring compared to mechanical abrasion, laser or spark-erosion, and as a result, it is a slow process. Therefore coarse structuring was performed at a high ion flux (25-40~nA). It is important to note that high ion flux cutting does not generate more defects: Higher currents are achieved by using broader beams, so that the flux density of ions impacting on the sample remains low. Further, the beam is purposely defocused to spread out power over an even bigger beam spot. Cutting a crystal into the shape shown in Fig.~1 takes about 12h.

The broad beam spot used for coarse cutting leads to rough and rounded edges. Therefore it is essential to polish the sidewalls. The polishing procedure involves cutting the last micron using smaller currents with smaller spot sizes. Three currents of 2.5~nA, 780~pA and 320~pA were used successively as the sidewall approached the final desired thickness. 

{ \bf Width determination:} The presented results critically rely on a precise determination of the effective width of the sample. The milling process will always create slightly canted sidewalls. After an initial rounding on the top, basically vertical sidewalls were achieved. The effective width was then calculated by taking the average as (2 $W_{bottom}$+$W_{top}$)/3. This weighted average takes the initial rounding of the edges at the top into account. An example measurement using Scanning Electron Microscope (SEM) images is given in Figure S2. Most importantly, the same methodology was used throughout the study to ensure consistent determinations of the sample width. 

{\bf Surface damage:} Another important aspect concerns the nature and depth of the ion beam induced surface damage layer. While the strong Shubnikov-de Haas oscillations observed in our patterned samples clearly evidence the high crystal quality of the bulk, a surface damage layer of a priori unknown extent surrounds the pristine core. To estimate the effective thickness of the amorphous layer, we performed a full damage cascade Monte Carlo simulation using the widely used software SRIM (Stopping and Range of Ions in Matter) ~\textit{(36)}.

Fig. S3 shows the depth profile of 30kV Ga ions impacting on the surface under realistic conditions of quasi-grazing incidence. The typical implantation profile follows a Gaussian distribution centered around the lateral stopping range, which is around 2~nm at this high incidence angle. The Ga implantation is suppressed exponentially by simple statistical arguments of a random walk by one decade every 6nm up to 20~nm, where a sudden drop in ion penetration signals a cut-off. At the same time, inelastic processes due to the ion-matter interaction generate phonons as well as defect cascades. These phenomena generally follow the profile of the implantation, which an integration over the inelastic processes in the simulation confirms. Therefore we estimate the damage layer to be on the order of 20~nm at each boundary in PdCoO$_2$.  We note that the relatively large atomic number of Pd is advantageous in reducing the ion implantation lengths.  SRIM calculations have been confirmed to accurately capture the Ga penetration during the FIB process, for example via atom probe tomography~\textit{(37)}, which also confirms the random nature of ion implantation.  This in turn ensures that the edges are rough from the point of view of electron boundary scattering.  This rough wall approximation is therefore adopted in our boundary scattering calculations. Combining the slight edge damage with our estimate of small lateral width variations after final polishing, we believe that a conservative estimate of the total error in our determination of width is 80 nm or less.  This would be approximately a 10$\%$ effect in our narrowest channel, falling successively for larger widths.

A check both on our width determination procedure and the overall width uncertainty is to compare the width obtained from the procedure outlined above with that deduced from magnetoresistance.  As discussed in the main manuscript and shown in Fig.~2, we observe well-defined maxima in the magnetoresistance for all studied widths.  The appearance of these maxima is a well-known phenomenon observed in very clean metals confined into structures smaller than the mean free path. The existence of a well-defined maximum evidences that a reasonably well-defined ``effective width" exists (a triangular cross-section would not result in such a maximum).  Further, the field values of the maxima for channel widths greater than 2 $\mu$m follow the relationship $\textit{B}_{max} = 0.62\frac{\hbar\textit{k}_F}{\textit{We}}$  to high accuracy.  If, instead of relying on the width measurements, we use the measured field maxima to deduce the widths of our narrowest channels, we see excellent agreement between the two methods: 0.73 $\mu$m (SEM measurement) cf 0.79 $\mu$m (magnetoresistance peak);  1.17~$\mu$m (SEM measurement) cf 1.15~$\mu$m (magnetoresistance peak);  1.90~$\mu$m (SEM measurement) cf 2.09~$\mu$m (magnetoresistance peak).  This gives us further confidence both in the method we have used to deduce the width from SEM images and in our error estimates. 

\subsection{Checks for length dependence and possible internal heating. }

When studying transport properties on short length scales, there is the possibility of creating significant Joule heating resulting in both a raised average electronic temperature and a temperature distribution across the sample leading to a position-dependent resistivity.  If these effects were very large, they might also lead to non-linear I-V characteristics, especially in the presence of phonon drag.  We verified during our main experiment that we resolved no I-V non-linearity, but since this would be a second-order effect, that does not necessarily rule out significant average sample heating.

Before discussing explicit tests of the electronic temperature and its distribution, we note that a large effect is unlikely.  Although current-dependent heating is an issue in samples with high resistances, our experiments are performed on an extremely good metal with a carrier concentration a factor 10$^4$ higher than those in the semiconductor devices studied in the beautiful work of the 1990s~\textit{(3, 4)} .  Further, we have approximately 16000 layers in parallel when we work at our standard constant r.m.s. measurement current of 10~mA, 177~Hz (corresponding to voltages in the range 200~nV to 2~mV r.m.s depending on the device dimensions).  For the narrowest channel we reduced this to 8000 layers in parallel, but worked at 1~mA r.m.s current.  This means that we are applying a maximum of just over 0.5~$\mu$A r.m.s. per layer.  Typical power dissipation in our devices is in the range nano- to microwatts, several orders of magnitude lower than the cooling power of the $^4$He cryostat used for the measurements, and the sample is thermally anchored to the external thermal reservoir through high conductivity metallic current and voltage leads and through direct epoxy contact along its length.
That these methods of thermal sinking are effective can be seen in several ways.  Firstly, we observed high resolution quantum oscillations in the resistivity when working at a reservoir temperature of 1.9~K.  The oscillations have a strong intrinsic temperature dependence when observed in equilibrium magnetic measurements ~\textit{(18)}, and the spectrum we saw (Fig.~3C) would not have been observable if the average electron temperature had been even one degree higher.  Further, we saw no change to the relative weighting of the frequencies as the power dissipation in the device changed by over a factor of two during the experiment.  Both of these observations are consistent with the electronic temperature being very close to that of the reservoir at our measurement currents.
Further checks on this come from the length dependence of the measured resistances, which scaled linearly with length in all studied devices, even up to the largest meandered device used for the quantum oscillation experiments (Fig.~1B). With a total length of 4600~$\mu$m, this is the largest structure we studied. The thickness of crystals of this macroscopic size is never homogeneous over such large distances, and indeed we observe steps in the thickness (Fig.~1B). Nonetheless we can calculate the average dimensions of the device: Length 4600~$\mu$m, thickness 17.1~$\mu$m, width 6~$\mu$m. In zero applied field, this device has a 4-terminal resistance of 9~m$\Omega$ at 2~K, and using these average dimensions we obtain a resistivity of 20~n$\Omega$ cm. We conservatively estimate an error of about 15\% for the uncertainties in the geometry. In spite of being 46 times longer than the device used for the channel narrowing experiment, the two resistivities agree for $\it{W}$ = 6~$\mu$m to within experimental error. 
To perform a still more precise check for length scaling of the resistance within the same crystal, and also to check for the influence of the length of the current path, we fabricated the device shown in Fig.~S4.  The deviations in resistivity between the accessible equally spaced voltage leads (50~$\mu$m spacing) scales with length as expected within an experimental uncertainty of 2\%. For example, sourcing a current through the main structure (colored purple in Fig.~S1) at 2~K, we measure a resistance R$_{12}$ = 89.7~m$\Omega$ between the contacts V$_{1}$ and V$_{2}$; and R$_{13}$ = 177.9~m$\Omega$  between V$_{1}$ and V$_{3}$. The ratio R$_{13}$ / R$_{12}$ = 1.983 agrees well with the ratio of the device length, L$_{13}$ / L$_{12}$ = 100~$\mu$m / 50~$\mu$m = 2 as expected for the usual linear dependence of resistance to conductor length. 

In this structure, we can also directly check for influence of the current path length. The total length of the conductor between the main current pads is 2700~$\mu$m  (purple). By injecting the current alternatively through the contacts V$_{1}$ and V$_{4}$, the effective length of the current path is reduced by more than a factor of 3 to 840~$\mu$m . The resistance R$_{23}$ measured between the central voltage contacts V$_{2}$ and V$_{3}$ e is the same within 2\% accuracy for both current cases (89.7~m$\Omega$  at 2700~$\mu$m  conductor length, 91.2~m$\Omega$ at 840~$\mu$m  conductor length). 

Taken together, we believe that these checks rule out a significant influence of Joule heating or other sources of systematic error leading to length-dependent resistivity in our experiments.

\section{Hydrodynamic electrons: theory}

We first shall give an overview of the theoretical description of
the electrons as a two-dimensional hydrodynamic fluid of charged
particles obeying Fermi statistics.  We assume that the scattering of
the electrons is comprised of three components, namely
momentum-relaxing collisions with impurities and wire boundaries and
momentum-conserving electron-electron scattering.  In particular we
ignore electron-phonon interactions as well as electron-electron
umklapp scattering processes.  Our calculation closely follows the
beautiful work of de Jong and Molenkamp~\textit{(4)} and
reproduces their result.  Secondly we illustrate a simple hydrodynamic
model describing the electrons as a flow of classical charged
particles characterized by their viscosity.

\subsection{Momentum conserving and momentum relaxing scattering}
\label{sec:dejong}

Following Ref.~\textit{(4)}, we start with a semiclassical
description of the motion of the electrons, and use the standard
Boltzmann transport equation
\begin{equation}
    \frac{{\rm d}f}{{\rm d}t}=
    \frac{\partial f}{\partial t}
    +
    \sum_{i}\left(
    \frac{\partial x_{i}}{\partial t}
    \frac{\partial f}{\partial x_{i}}
    +
    \frac{\partial v_{i}}{\partial t}
    \frac{\partial f}{\partial v_{i}}
    \right)
    =
    \left.\frac{\partial f}{\partial t}\right|_{\text{imp}}
    \label{eqn:boltzmanntransport}
\end{equation}
for the distribution function $f(\vec x,\vec v)$ in the phase space of
the electrons at positions $\vec x=(x_{1},x_{2})^{T}$ with velocity
$\vec v=(v_{1},v_{2})^{T}$ in a two-dimensional wire.  The term on the
right-hand side of the equation denotes the momentum relaxing
collisions of the electrons violating Liouville's theorem.

The electrons with mass $m$ are subject to a Lorentz force $\vec
F=\hbar(\partial\vec k/\partial t)=m(\partial\vec v/\partial t)=-e\vec
E$ in the applied static electric field, thus we can replace 
Eq.~(\ref{eqn:boltzmanntransport}) with
\begin{equation}
    \sum_{i}\left(
    v_{i}\frac{\partial f}{\partial x_{i}}
    -\frac e\hbar E_{i}\frac{\partial f}{\partial k_{i}}
    \right)
    =
    \left.\frac{\partial f}{\partial t}\right|_{\text{scatt}},
    \label{eqn:boltzmann}
\end{equation}
where we define $(\partial f/\partial t)_{\text{scatt}}$
to formally comprise all remaining contributions to $f$.

At equilibrium, $f$ is given by the Fermi-Dirac distribution
$f_{0}=\left[\exp\left(\beta(\epsilon-\mu)\right)+1\right]^{-1}$ with
single-particle energies $\epsilon=\hbar^{2}/(2m)\vec k^{2}=(m/2)\vec
v^{2}$ and Fermi energy $\mu$.  The temperature of the electrons in
the wire is $T=(k_{\text B}\beta)^{-1}$.  At small electric fields
applied in $x_{1}$ direction along the wire, we can expand around 
the equilibrium,
\begin{equation}
    f(\vec x,\vec v)=f_{0}+
    \left(-\frac{\partial f_{0}}{\partial\epsilon}\right)
    \chi(x_{2},\phi),
    \label{eqn:lb}
\end{equation}
which implies that the nonequilibrium part of the distribution
function in momentum space is only in a small area around the
(circular) equilibrium Fermi surface.  Spatially $f(\vec x,\vec v)$
only depends on the transverse coordinate $x_{2}$, and its velocity
dependence, writing $\vec v=v\hat v$ with $\hat
v:=(\cos\phi,\sin\phi)^{T}$, is split into an energy dependent part
given by $(\partial f_{0}/\partial\epsilon)$ and an explicit directional
dependence parametrized by the angle $\phi$ with respect to the
$v_{1}$ direction.  With this distribution function the current
density can be evaluated according to
\begin{eqnarray}
    \vec\jmath(x_{2})
    &=&
    2e\int{\rm d}^{2}vf(\vec x,\vec v)\vec v
    \nonumber\\
    &=&
    e\int{\rm d}\epsilon{\cal D}(\epsilon)
    \left(-\frac{\partial f_{0}}{\partial\epsilon}\right)
    \int_{0}^{2\pi}\frac{{\rm d}\phi}{2\pi}
    \chi(x_{2},\phi)\vec v
    \nonumber\\
    &=&
    \frac{e{\cal D}v_{\text F}}{2\pi}\int_{0}^{2\pi}{\rm d}\phi
    \chi(x_{2},\phi)\hat v.
    \label{eqn:j}
\end{eqnarray}
We are using $\vec v\approx\vec v_{\text F}=v_{\text F}\hat v$ and a
constant density of states ${\cal D}(\epsilon)=m/(\pi\hbar^{2})$.

Inserting Eq.~(\ref{eqn:lb}) into the
Boltzmann equation~(\ref{eqn:boltzmann}) gives to linear order
\begin{equation}
    v_{2}\frac{\partial\chi(x_{2},\phi)}{\partial x_{2}}-eEv_{1}=
    \left.
    \frac{\partial\chi(x_{2},\phi)}{\partial t}
    \right|_{\text{scatt}}
    \label{eqn:chi}
\end{equation}
as the determining equation for the unknown $\chi(x_{2},\phi)$.

Following Ref.~\textit{(4)}, we make a relaxation-time 
approximation for the bulk impurity scattering part,
\begin{equation}
    \left.\frac{\partial\chi(x_{2},\phi)}{\partial t}\right|_{\text{MR}}
    =
    -\frac{\chi(x_{2},\phi)}{\tau_{\text{MR}}}.
    \label{eqn:chib}
\end{equation}
The momentum-conserving electron-electron scattering part is
parametrized as
\begin{eqnarray}
    \lefteqn{\left.
    \frac{\partial\chi(x_{2},\phi)}{\partial t}
    \right|_{\text{MC}}=}
    \label{eqn:chimc}\\&&
    -\frac{\chi(x_{2},\phi)}{\tau_{\text{MC}}}
    +\frac1{2\pi\tau_{\text{MC}}}
    \int_{0}^{2\pi}{\rm d}\phi'\chi(x_{2},\phi')
    \left(1+2\hat v^{T}\hat v'\right),
    \nonumber
\end{eqnarray}
which is the most simple momentum-conserving form for the scattering
term assuming that the electrons relax to a shifted Fermi-Dirac
distribution $f(\vec x,\vec v)\approx f_{0}(\epsilon-m\vec v^{T}\vec 
v_{\text{drift}})$ where the drift velocity is related to the current 
density via $\vec\jmath(x_{2})=ne\vec v_{\text{drift}}(x_{2})$, $n:={\cal 
D}\mu$ being the electron density.

For the boundary scattering we assume diffusive reflection.  Given a
wire of width $W$, this requires for the solution of 
Eq.~(\ref{eqn:lb}) that
\begin{eqnarray}
    \chi(-W/2,\phi)
    &=&
    \frac1\pi\int_{\pi}^{2\pi}{\rm d}\phi'\chi(-W/2,\phi'),
    \\
    \chi(W/2,\phi)
    &=&
    \frac1\pi\int_{0}^{\pi}{\rm d}\phi'\chi(W/2,\phi')
\end{eqnarray}
at the transverse boundaries of the wire.  (We note that
$\phi\in[0,\pi]$ for $x_{2}=-W/2$ and $\phi\in[\pi,2\pi]$ for
$x_{2}=W/2$.)

We introduce an effective mean free path
$\ell_{\text{eff}}(x_{2},\phi)$ describing the average length an
electron at position $x_{2}$ travels in the direction given by the
angle $\phi$ after its last momentum-relaxing scattering event 
through the parametrization
\begin{equation}
    \chi(x_{2},\phi)=eE\cos(\phi)\tilde\ell_{\text{eff}}(x_{2},\phi).
    \label{eqn:chiparam}
\end{equation}
From Eqs.~(\ref{eqn:j}) and~(\ref{eqn:chiparam}) it follows that its 
angular average
\begin{equation}
    \ell_{\text{eff}}(x_{2})
    :=
    \frac1\pi\int_{0}^{2\pi}{\rm d}\phi
    \cos^{2}(\phi)\tilde\ell_{\text{eff}}(x_{2},\phi)
\end{equation}
is proportional to the drift velocity,
\begin{equation}
    \vec v_{\text{drift}}(x_{2})
    =
    \frac{e\vec E}{mv_{\text F}}\ell_{\text{eff}}(x_{2}).
\end{equation}
Thus defining
\begin{equation}
    {\cal L}_{\text{eff}}:=\frac1W\int_{-W/2}^{W/2}{\rm d}x_{2}
    \ell_{\text{eff}}(x_{2})
\end{equation}
the conductivity of the wire is given by
\begin{equation}
    \sigma=\frac{ne^{2}}{mv_{\text F}}{\cal L}_{\text{eff}}.
\end{equation}

Together with the approximations~(\ref{eqn:chib})
and~(\ref{eqn:chimc}) we insert Eq.~(\ref{eqn:chiparam}) into
Eq.~(\ref{eqn:chi}) and transform the resulting differential 
equation into a Fredholm integral equation of the second kind,
\begin{eqnarray}
    \ell_{\text{eff}}(x_{2})
    &=&
    \hat\ell_{\text{eff}}(x_{2})
    \nonumber\\&&{}
    +\lambda
    \int_{-W/2}^{W/2}{\rm d}x_{2}'K(x_{2},x_{2}')
    \ell_{\text{eff}}(x_{2}'),
    \label{eqn:leff}
    \\
    \hat\ell_{\text{eff}}(x_{2})
    :&=&
    \ell-\frac{2\ell}{\pi}\int_{0}^{\pi/2}{\rm d}\phi
    \cos^{2}(\phi)
    \left(
    {\rm e}^{-(W/2+x_{2})/(\ell\sin\phi)}
    \right.\nonumber\\&&\left.{}
    +{\rm e}^{-(W/2-x_{2})/(\ell\sin\phi)}
    \right),
    \\
    K(x_{2},x_{2}')
    &:=&
    \frac1\pi\int_{0}^{\pi/2}{\rm d}\phi\frac{\cos^{2}\phi}{\sin\phi}
    {\rm e}^{-|x_{2}-x_{2}'|/(\ell\sin\phi)}
\end{eqnarray}
with $\lambda:=1/\ell_{\text{MC}}$.  Here we have introduced the
``bare'' mean free path $\ell$ with
$1/\ell:=1/\ell_{\text{MR}}+1/\ell_{\text{MC}}$ and
$\ell_{\text{MC}}:=v_{\text F}\tau_{\text{MC}}$,
$\ell_{\text{MR}}:=v_{\text F}\tau_{\text{MR}}$.

\subsection{Numerics}
\label{sec:numerics}

For given values of the momentum conserving and the momentum relaxing
mean free paths, together with the width of the wire, we eventually
solve Eq.~(\ref{eqn:leff}) numerically: Measuring lengths in units of
$W$ we discretize $x_{2}$ with $N\le100$ segments at 
$\{x_{2}=x_{j}:j=1\ldots N\}$ with widths
$\{w_{j}:j=1\ldots N\}$ between the $\pm W/2$ boundaries and replace
the integration over the kernel $K$ in each of the $N$ partitions by a
mean value, mapping the solution onto a matrix problem, written in
components
\begin{equation}
    \ell_{i}-\lambda\sum_{j=1}^{N}K_{ij}w_{j}\ell_{j}
    =
    \hat\ell_{i},
    \quad
    i=1\ldots N
    \label{eqn:matrix}
\end{equation}
with $K_{ij}:=K(x_{i},x_{j})$ and equivalently for $\ell$ and
$\hat\ell$.  The solutions to this linear system can be found easily
as long as $\lambda\equiv(\ell_{\text{MC}}/W)^{-1}$ is not too large,
i.\,e.  for weak electron-electron scattering.  In the opposite case
$\ell_{\text{MC}}/W\ll1$, Eqs.~(\ref{eqn:leff}) and~(\ref{eqn:matrix})
become numerically unstable.  To overcome this instability, we solve
the equivalent problem
\begin{eqnarray}
    \lefteqn{\ell_{\text{eff}}(x_{2})\left(
    1-\lambda\int_{-W/2}^{W/2}{\rm d}x_{2}'
    K(x_{2},x_{2}')\right)=}
    \\
    &&
    \hat\ell_{\text{eff}}(x_{2})
    +
    \lambda
    \int_{-W/2}^{W/2}{\rm d}x_{2}'K(x_{2},x_{2}')
    \left(
    \ell_{\text{eff}}(x_{2}')
    -\ell_{\text{eff}}(x_{2})
    \right),
    \nonumber
\end{eqnarray}
again mapping it onto a matrix problem as before.
Fig.~S5 displays the resulting dependence of the
resistivity $\rho$ on the ratio $\ell_{\text{MR}}/W$ for fixed values
$\ell_{\text{MC}}/\ell_{\text{MR}}$.

\subsection{Viscous flow of charged particles}

In this section, we present a calculation of the flow of charged
particles through a two-dimensional channel in the fully hydrodynamic
limit, i.\,e. when momentum is fully conserved in the bulk of the fluid.
We start from Newton's second law $\rho\dot{\vec v}=\vec g$,
introducing a mass density $\rho:=m\int{\rm d}^{2}vf(\vec x,\vec v)$
and a force field $\vec g(\vec x)$.  The electrons are subject to two
forces, (a) shear forces characterized by a finite viscosity, and (b)
an electrostatic field gradient along the wire in direction $x_{1}$
(Lorentz force).  We regard the electrons as an incompressible fluid
and ignore all momentum relaxing processes.  In particular we assume
that no point scatterers are present in the wire.

The shear modulus of an infinitesimally small cube inside the fluid is
\begin{equation}
    S_{ij}:=\eta\left(\frac{\partial v_{i}}{\partial x_{j}}+
    \frac{\partial v_{j}}{\partial x_{i}}\right),
\end{equation}
and the $i$th component of the corresponding force field thus is
\begin{eqnarray}
    g_{i}
    &=&
    \sum_{j}\frac{\partial S_{ij}}{\partial x_{j}}
    =
    \eta\left[
    \left(\sum_{j}\frac{\partial^{2}}{\partial x_{j}}\right)v_{i}
    +\frac{\partial}{\partial x_{i}}
    \left(\sum_{j}\frac{\partial v_{j}}{\partial x_{j}}\right)
    \right],
    \nonumber
\end{eqnarray}
where the second term inside the square brackets vanishes due to the
incompressibility of the electrons ($\nabla\vec v\equiv0$).  Together
with the electric field $\vec E$ along the wire we have to solve the 
equation
\begin{equation}
    \rho\dot{\vec v}=\eta\Delta\vec v-\rho\frac em\vec E,
\end{equation}
which is equivalent to the Navier-Stokes equation for a stationary
laminar flow of an incompressible fluid.  We seek a steady state, so
$\dot{\vec v}=0$ as well.  With $\vec E=-(V/L)\hat x_{1}$ for our wire
with length $L$ and width $W$ ($\hat x_{i}:=x_{i}/|\vec x|$), the
solution is given by $\vec v=v(x_{2})\hat x_{1}$,
\begin{equation}
    v(x_{2})=\frac12\frac\rho\eta\frac emE
    \left(x_{2}^{2}-\frac{W^{2}}{4}\right)
\end{equation}
for $-W/2\le x_{2}\le W/2$.  With
$n=\rho/m$, this causes a current
\begin{equation}
    I
    =
    neT\int_{-W/2}^{W/2}{\rm d}x_{2}v(x_{2})
\end{equation}
through the wire of thickness $T$ (being infinitely smooth at
$x_{3}=\pm T/2$), and we obtain for the resistance $R=V/I$
\begin{equation}
    R=\frac{12 L}T\frac{\eta}{(ne)^{2}}
    \frac1{W^{3}}.
    \label{eqn:r}
\end{equation}
We note that the finiteness of the resistance is exclusively due to
the finite width of the wire: Because the electronic momentum is
conserved, we must have $R\to0$ in the bulk limit.

\subsection{Intuitive significance of the de Jong-Molenkamp theory}

The comparison of Eq.~(\ref{eqn:r}) with the solid line in
Fig.~S5 is significant.  Resistance $R$ varying as $1/W^3$
is equivalent to resistivity $\rho$ varying as $1/W^2$ because of the
extra geometrical factor $TW/L$ in the definition of $\rho$.  The
quadratic extra contribution to $\rho$ in the calculation of
sections~\ref{sec:dejong} and~\ref{sec:numerics} at high rates of
momentum-conserving scattering
($\ell_{\text{MC}}/\ell_{\text{MR}}\to0$) is therefore naturally
identified with a viscous contribution to the resistivity.  In the
opposite limit when the rate of momentum-conserving scattering tends
to zero ($\ell_{\text{MC}}/\ell_{\text{MR}}\to\infty$) the dotted line
limits to the ballistic transport result calculated in theories that
ignore momentum-conserving scattering~\textit{(30)}. Since the
theory limits to physically reasonable results at both its extremes,
we believe that some confidence can be placed in its predictions.
When plotted in the dimensionless units of Fig.~S5, these
predictions are unique for each value of
$\ell_{\text{MC}}/\ell_{\text{MR}}$, and contain no free parameters.
In other words, both the functional form and the magnitude of the
change in resistivity as a function of channel have physical
significance.  In this context, the closeness of the prediction for to
the experimental data shown in Fig.~4A of the main paper for
$\ell_{\text{MC}}/\ell_{\text{MR}}=0.1$ is very good.

The $1/W^3$ variation predicted for $R$ at
$\ell_{\text{MC}}/\ell_{\text{MR}}=0.005$ gives the opportunity to
calibrate $\rho/\rho_0-1$ against viscosity.  Specifically, for 
$\ell_{\text{MC}}/\ell_{\text{MR}}\ll1$, we approximately have
\begin{equation}
    \frac{\rho}{\rho_{0}}
    \approx
    1+b\left(\frac{\ell_{\text{MR}}}{W}\right)^{2},
    \label{eqn:rapprox}
\end{equation}
with a coefficient $b$ strongly depending on the exact
$\ell_{\text{MC}}/\ell_{\text{MR}}$ ratio.  We note that in the strong
electron-electron scattering limit we are at the border of validity of
the theory, mirrored by the fact that we suffer from numerical
instabilities when solving Eq.~(\ref{eqn:leff}) for
$\ell_{\text{MC}}/\ell_{\text{MR}}\lesssim0.005$.  With these
ambiguities in mind, we obtain an estimate $b={\cal
O}(10^{-3}\ldots10^{-2})$.

\section{Analysis of experimental data using hydrodynamic theory}

\subsection{Estimate of viscosity}
For analyzing the actual experimental data from PdCoO$_{\text2}$, we
first note that the momentum conserving scattering processes relevant
to our experiment likely include normal electron-phonon events (since
the strong phonon drag prevents these from relaxing the momentum of
the electron assembly) as well as electron-electron events.  However,
both types of processes can safely be encoded by the parameter
$\ell_{\text{MC}}$ for the purposes of our analysis, since it is
performed at fixed temperature.  A more sophisticated theory would be
required to extend it reliably to situations in which the temperature
is varied.

Next, we estimate the viscosity $\eta$.  First we note that 
comparison of Eqs.~(\ref{eqn:r}) and~(\ref{eqn:rapprox}) implies that
\begin{equation}
    b
    =
    \frac{12\eta}{(ne)^{2}\rho_{0}\ell_{\text{mr}}^{2}}.
\end{equation}
Taking the estimate of the value of $b$ obtained from here, using the
known values for $n$, $e$, $\ell_{\text{MR}}$ and $\rho_{0}$
($4.9\times10^{28}\,\rm m^{-3}$, $1.602\times10^{-19}\,\rm C$,
$2\times10^{-5}\,\rm m$ and $8\times10^{-11}\,\Omega\rm m$
respectively) and working at our maximum measured value of
$\ell_{\text{MR}}/W=20$, for which $\rho/\rho_{0}-1\approx4$ for
$\ell_{\text{MC}}/\ell_{\text{MR}}=0.005$ (Fig.~S5, solid
line), we obtain $\eta$ in the range
$1.6\times10^{-4}\ldots1.6\times10^{-3}\,{\rm kg}({\rm m\,s})^{-1}$.
Converting this unambiguously to the viscous contribution to our
experimental data at $\ell_{\text{MR}}/W=20$ would require being able
to decouple the viscous effects on boundary scattering from those
associated with the impurity scattering.  This is possible for the
$\ell_{\text{MC}}/\ell_{\text{MR}}\to\infty$ limit (impurity
scattering dominates) and $\ell_{\text{MC}}/\ell_{\text{MR}}\to0$
(viscous effects dominate) but cannot be done with certainty for the
parameter range ($\ell_{\text{MC}}/\ell_{\text{MR}}\approx0.1$)
relevant to the experiments.  However, two limits can be established
on the viscous contribution to the experimental data.  One extreme is
to attribute the entire extra resistivity to viscous effects, i.e. to
set $(\rho/\rho_0-1)_{\text{visc}}=14$.  The other is to attribute the
viscous contribution only to the difference between the measured data
and the boundary scattering that would have resulted in the absence of
any momentum conserving scattering processes.  This sets
$(\rho/\rho_0-1)_{\text{visc}}=6$.  The true viscous contribution must
lie between these limits, yielding an estimate for the viscosity of
the electron fluid in PdCoO$_{\text{2}}$ of $\eta$ between
$3\times10^{-4}$ and $6\times10^{-3}\,{\rm kg}({\rm m\,s})^{-1}$.

Viscosities of everyday fluids are typically quoted either as dynamic viscosity $\eta$ as given above, or as kinematic viscosity $\nu$, obtained by dividing out the mass density.  Doing that for our PdCoO$_{\text{2}}$ results gives 0.01~m$^2$s$^{-1}$$<\nu<$ 0.3~m$^2$s$^{-1}$.  In a Fermi liquid, $\nu$  can be estimated as $\alpha\nu{_F}\ell_{MC}$ , with $\alpha \sim$0.2 \textit{(38)} .  Fig.~4 of the main paper show that our data are consistent with $\ell_{MC}$$\sim$2~$\mu$m, and $\nu{_F} $ for PdCoO$_{\text{2}}$ is approximately 7.5 $\times$ 10$^5$ ms$^{-1}$ ~\textit{(18)} , giving a second estimate for $\nu$ that falls within our quoted range.  This is a useful internal cross-check on our method for estimating viscosity.  Taken together, these analyses mean that the electron fluid in PdCoO$_{\text{2}}$ has a dynamic viscosity similar to that of water (1$\times$10$^3$~kg(ms)$^{-1}$ at room temperature) but a kinematic viscosity greater closer that of honey ($\sim$0.01~m$^2$s$^{-1}$  at room temperature).

After submission of our manuscript, a report of measurements of the viscosity of the electron fluid in graphene appeared on the archive~\textit{(39)}.  For those samples the kinematic viscosity is similar to that deduced here for PdCoO$_{\text{2}}$, but the dynamic viscosity is approximately 10$^4$ smaller due to the low carrier concentration.  

\subsection{Effect of varying $\ell_{MC}/\ell_{MR}$  by changing sample temperature}
As discussed in the main text, the theory of Ref.~\textit{(4)} cannot be expected to directly calculate temperature dependent hydrodynamic effects in PdCoO$_2$ accurately because it is likely to have different temperature dependent sources for its momentum-conserving scattering than those expected in the low carrier density two-dimensional electron gas for which the theory was designed.  However, one can anticipate that, by changing the temperature, one can alter the $\ell_{MC}/\ell_{MR}$ ratio.  Since the Fermi temperature of PdCoO$_2$ is so high (approximately 30000 K) that all thermally-induced momentum-relaxing scattering at room temperature and below is quasi-elastic, this is similar to studying the effects of changing residual resistivity, but with an unknown accompanying change to $\ell_{MC}$.  Although the $\ell_{MC}$ change is not controlled, such an experiment has the capability of checking for sensitivity to momentum-conserving scattering, as outlined in Figs.~S6 A and B.  In each figure, the $\ell_{MC}/\ell_{MR}$ ratio is changed by a factor of 20 by starting from different assumed values of $\ell_{MC}/\ell_{MR}$ and varying $\ell_{MR}$ .  In Fig. S6 A, where the starting value is 0.05 is in the region deduced from the analysis accompanying Fig.~4 C of the main paper, changing the ratio by decreasing $\ell_{MR}$ results in predicted changes to the data because the viscous contribution to the boundary scattering is being changed.  In Fig.~S6 B, in contrast, the only change resolved is restriction of the accessible $\ell_{MR}/W$ range.  This is because in this range of $\ell_{MC}/\ell_{MR}$ the momentum-conserving scattering is too weak to affect the resistance, and all the curves are essentially the same as the standard transport theory curve of Fig.~4 in the main paper.  

 In Fig.~S6 C we present analysis of data at 20, 30, 40, 50, and 75~K, a range across which  $\ell_{MR}$ changes substantially, by amounts that can be deduced from the restriction along the $\ell_{MR}/W$ axis.  The curves ``fan out'' from each other like the data predicted in Fig.~S6~A, rather than following the near-universal curve predicted in Fig.~S6~B or the completely universal curve that would be predicted by traditional transport theory.  We believe that this is further evidence that momentum-conserving scattering and hence electronic hydrodynamics play an important role in determining the resistance of our channels.

\pagebreak

\begin{figure}[t]
	\begin{center}
	\includegraphics[width=11 cm]{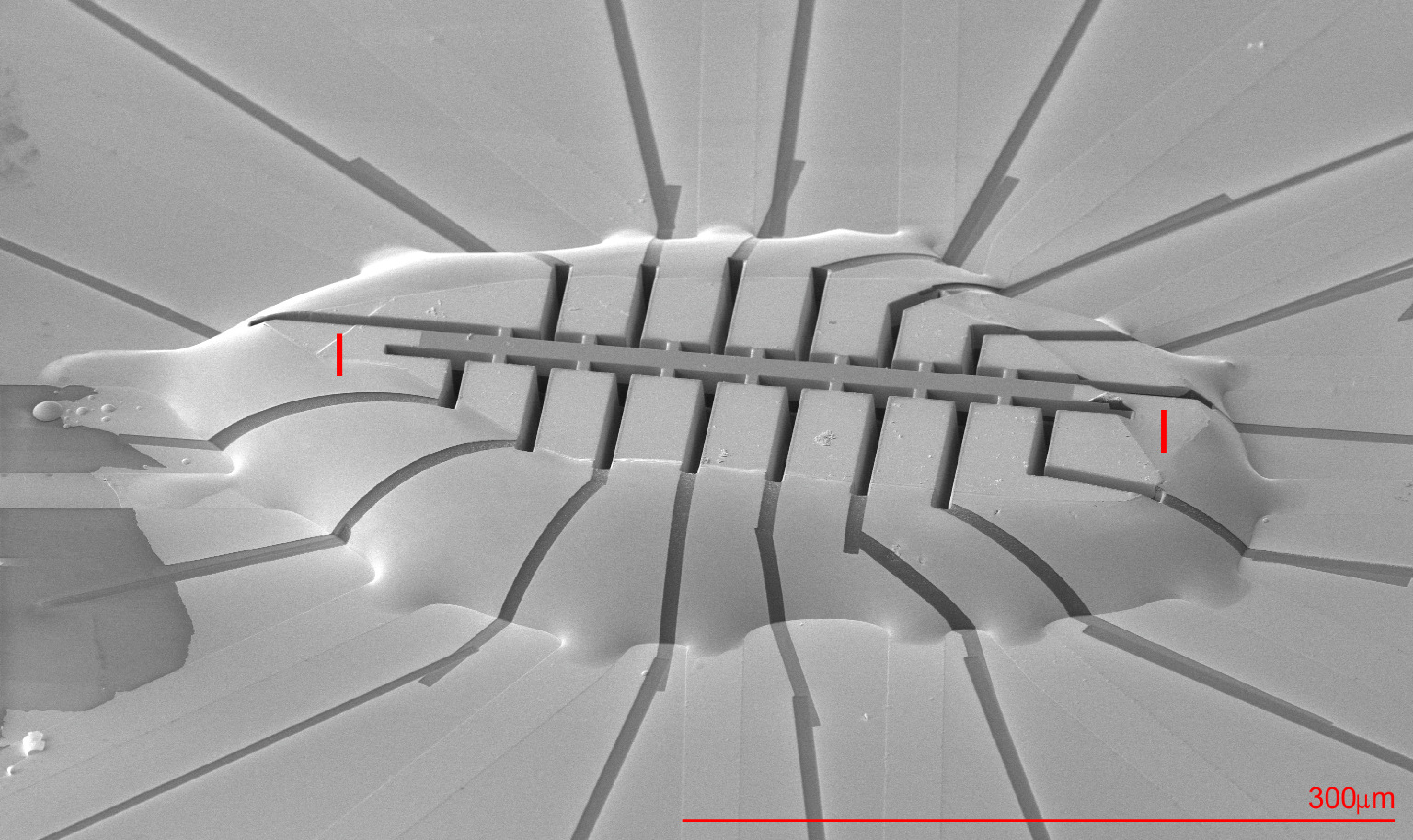}
	\end{center}
	\caption{\baselineskip1.5em\label{S1} The sample fabricated to check current flow and electrical thickness as a function of length along our device.  The voltage contacts are spaced with a separation approximately a factor of three closer together than on the device shown in Fig.~1 of the manuscript, allowing determination of the electrical thickness of that sample as described in the text. Current is injected through the top contacts marked `I'.}
\end{figure}

\begin{figure}[t]
	\begin{center}
	\includegraphics[width=8 cm]{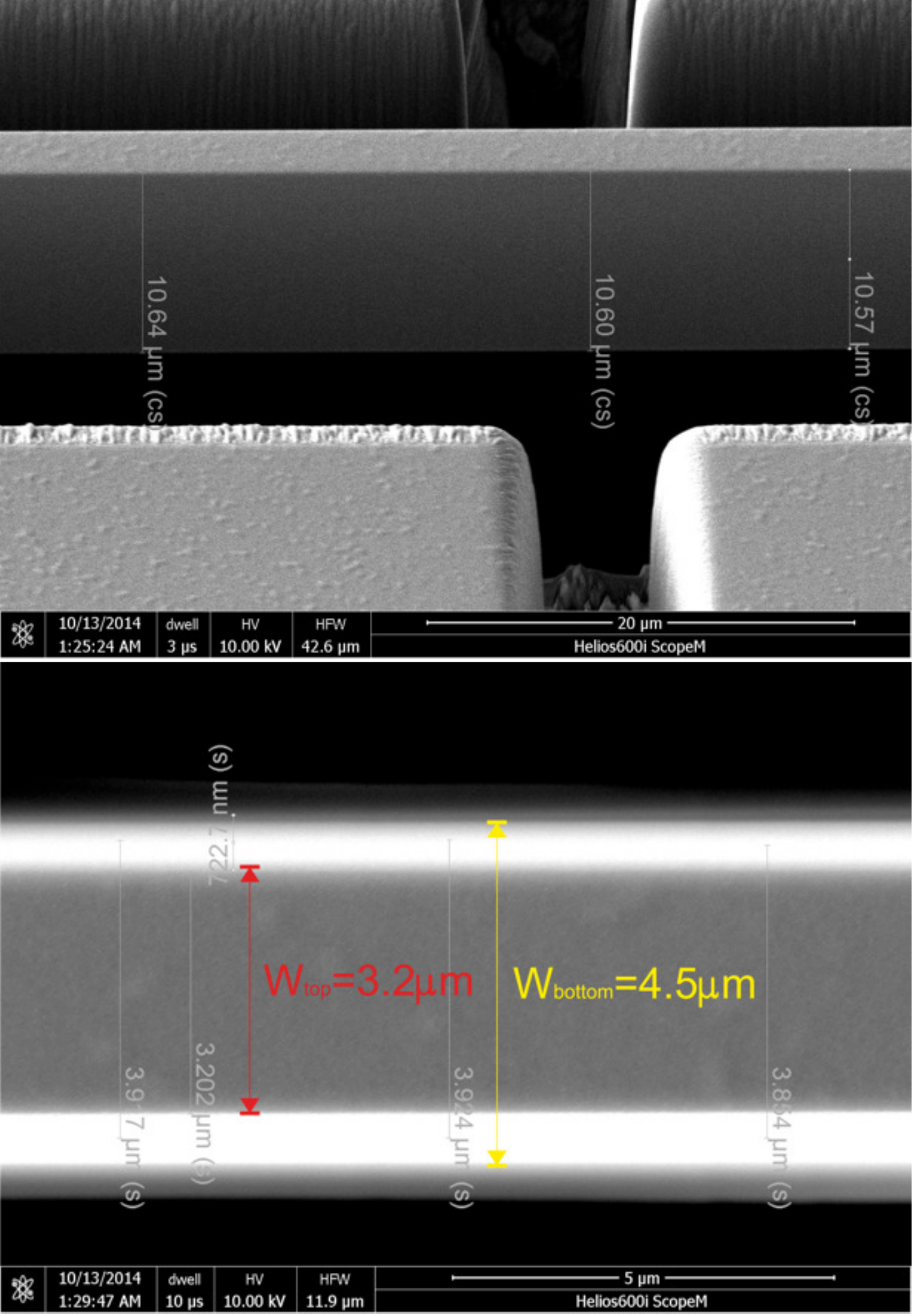}
	\end{center}
	\caption{\baselineskip1.5em\label{S2} Width and height determination in the SEM for a sample around $W$~=4~$\mu$m. In this case, $W_{top}$~=3.2~$\mu$m and $W_{bottom}$~=4.5~$\mu$m, yielding a weighted average width of $W$=4.06~$\mu$m.}
\end{figure}

\begin{figure}[t]
	\begin{center}
	\includegraphics[width=10 cm]{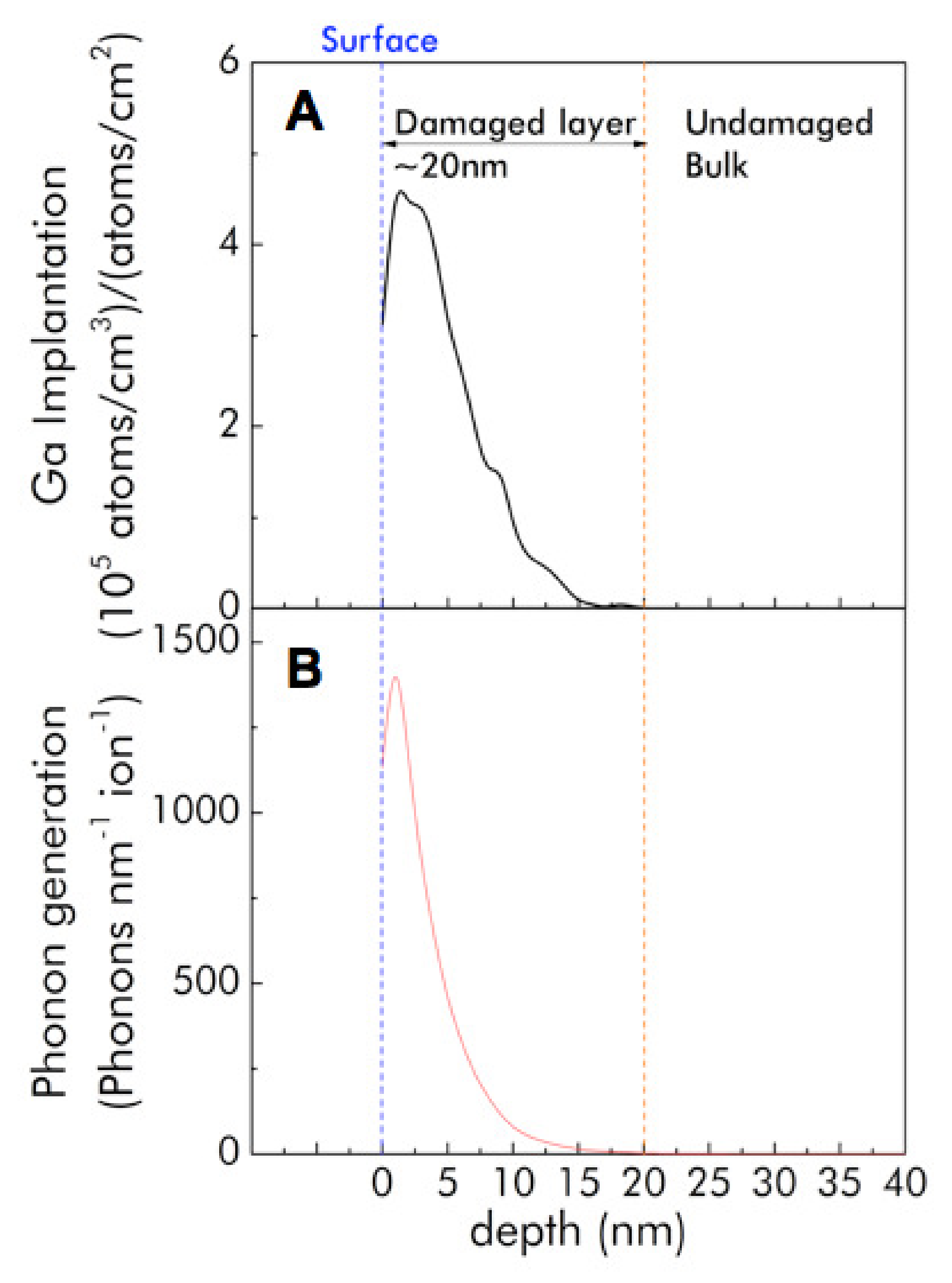}
	\end{center}
	\caption{\baselineskip1.5em\label{S3} Ga implantation and damage range in PdCoO$_2$ for high angle incidence (87.5$^o$).}
\end{figure}

\begin{figure}[t]
	\begin{center}
	\includegraphics[width=14 cm]{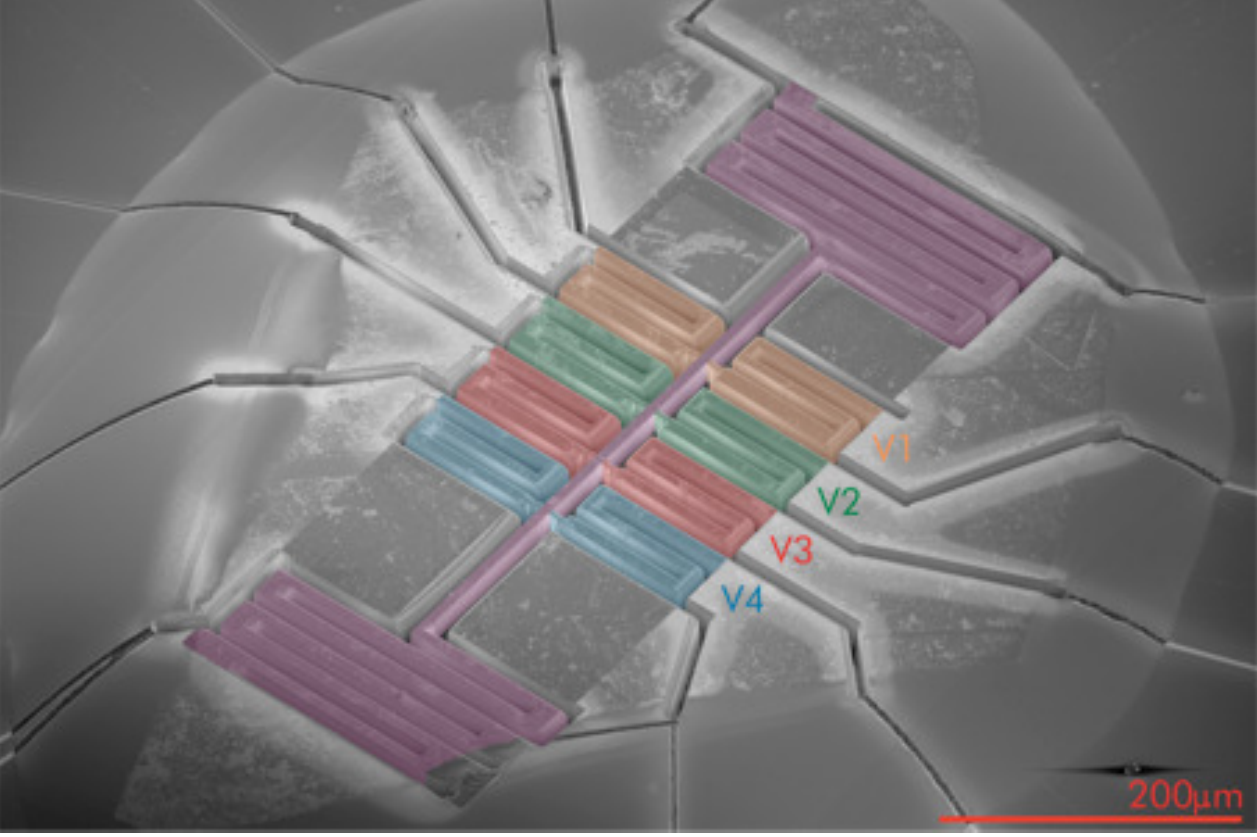}
	\end{center}
	\caption{\baselineskip1.5em\label{S4} PdCoO$_{2}$ microstructure designed for consistency checks of the resistivity scaling with conductor length. The main current path is highlighted in purple. The current is injected into the structure through large meandered paths, to ensure current homogeneity in the central bar.}
\end{figure}

\begin{figure}
    \centering
    \includegraphics[width=\columnwidth]{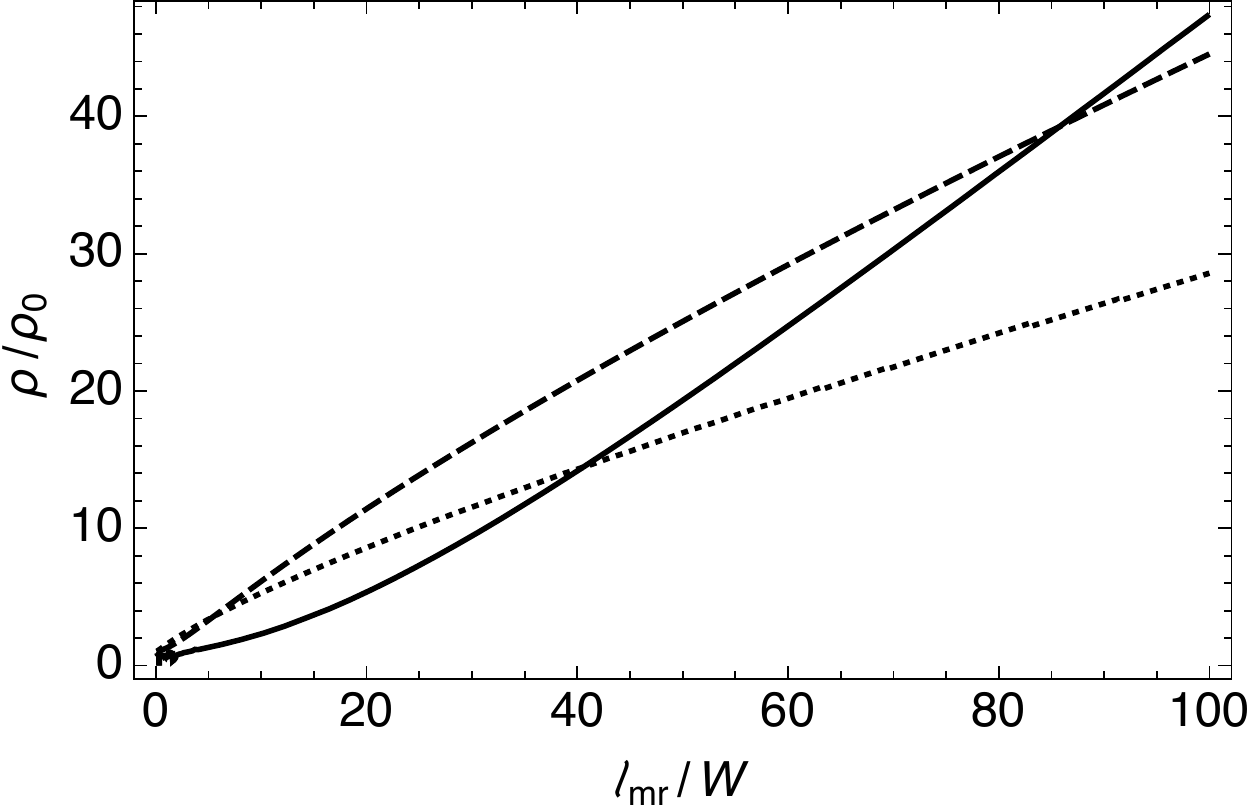}
    \caption{\baselineskip1.5em\label{S5} Dependence of the resistivity $\rho$ normalized to the
    bulk resistivity $\rho_{0}$ on the ratio $\ell_{\text{MR}}/W$ at
    fixed $\ell_{\text{MC}}/\ell_{\text{MR}}=0.005$ (solid line,
    strong electron-electron scattering), $0.1$ (dashed line), and
    $1000$ (dotted line, weak electron-electron scattering).}
    \label{fig:rho}
\end{figure}

\begin{figure}[t]
	\begin{center}
	\includegraphics[width=9.5 cm]{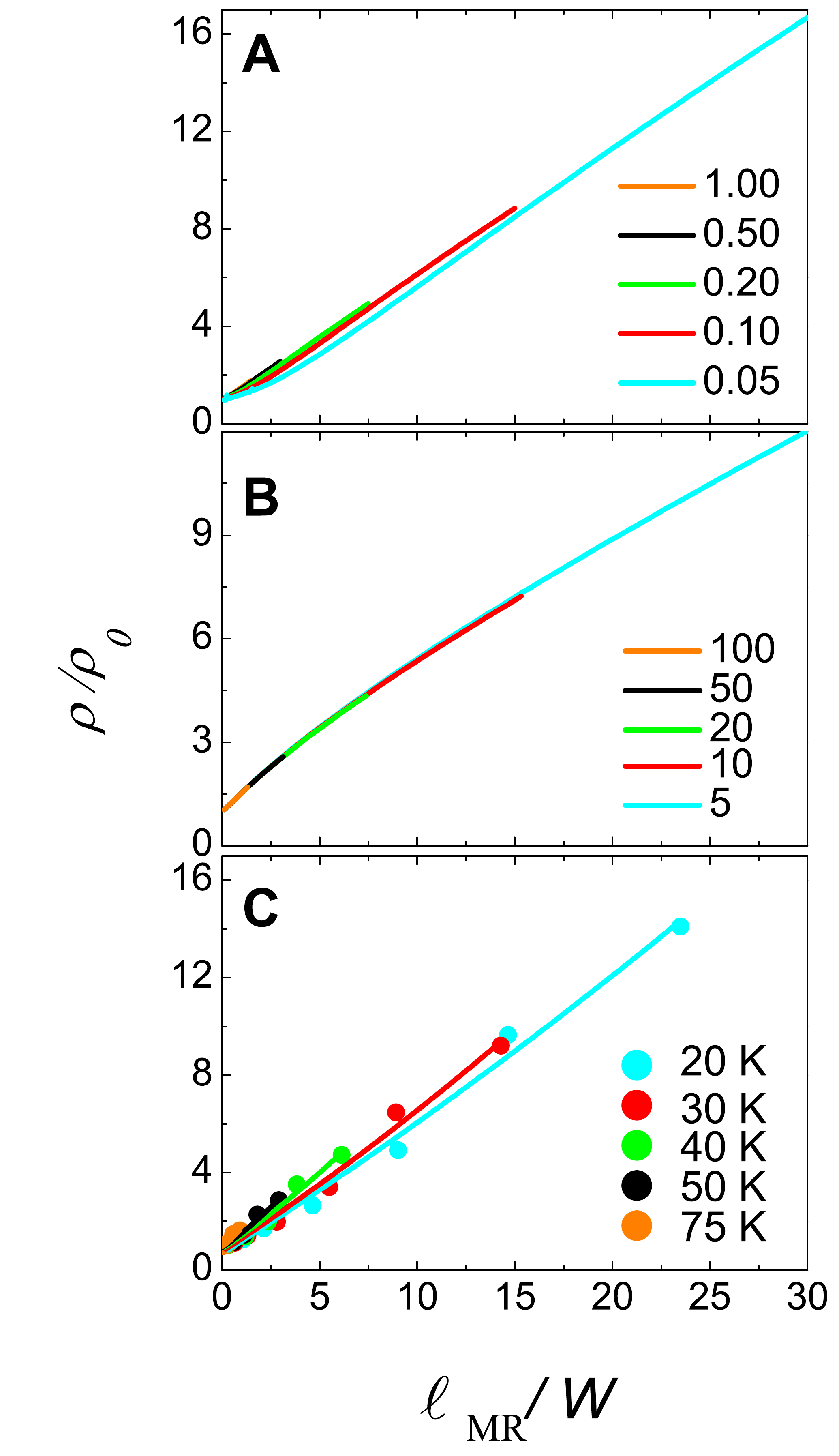}
	\end{center}
	\caption{ \baselineskip1.5em \label{S6} A and B:  Predictions of the hydrodynamic theory over similar dynamic ranges of $\ell_{MC}/\ell_{MR}$  but different starting values.  In A they are seen to fan out because in that part of the predicted phase diagram the results have sensitivity to changes in $\ell_{MC}/\ell_{MR}$ , while in B they follow a quasi-universal curve.  In C we show data from PdCoO$_2$ in which the $\ell_{MC}/\ell_{MR}$  ratio is tuned by raising the temperature. Lines are guides to the eye made using second-order polynomials, not fits to the theory. The precise changes to $\ell_{MC}/\ell_{MR}$ are not known, so the test is only qualitative, but the data are seen to be consistent with the prediction in panel A and definitely not consistent with the prediction in panel B.}
\end{figure}
\end{document}